\documentstyle[12pt,epsf,epsfig,cite]{article}

\newlength{\dinwidth}     
\newlength{\dinmargin}     
\def\lapproxeq{\lower .7ex\hbox{$\;\stackrel{\textstyle     
<}{\sim}\;$}}     
\def\gapproxeq{\lower .7ex\hbox{$\;\stackrel{\textstyle     
>}{\sim}\;$}}     
\def\be{\begin{equation}}     
\def\ee{\end{equation}}     
\def\bea{\begin{eqnarray}}     
\def\eea{\end{eqnarray}}     
     
\makeatletter     
\def\fmslash{\@ifnextchar[{\fmsl@sh}{\fmsl@sh[0mu]}}     
\def\fmsl@sh[#1]#2{%
\mathchoice     
{\@fmsl@sh\displaystyle{#1}{#2}}%
{\@fmsl@sh\textstyle{#1}{#2}}%
{\@fmsl@sh\scriptstyle{#1}{#2}}%
{\@fmsl@sh\scriptscriptstyle{#1}{#2}}}     
\def\@fmsl@sh#1#2#3{\m@th\ooalign{$\hfil#1\mkern#2/\hfil$\crcr$#1     
#3$}}     
\makeatother     
     
\begin{document}     
\begin{flushright}     
 July 4, 1998 \\     
BA-98-33\\
\end{flushright}     
     
\begin{center}     
\vspace*{2cm}     
{\Large \bf Phenomenological description of the $\gamma^\star p$ cross
section\\[3mm] at low $Q^2$}     
\\     
\vspace*{1cm}     
A.\ Rostovtsev$^1$, M.G.\ Ryskin$^2$, and R.\ Engel$^3$\\
\end{center}     
     
\vspace*{0.5cm}     
     
\begin{tabbing}     
$^1$xxxxx\= \kill     
      
\indent $^1$ \> ITEP, B.\ Cheremushkinskaja 25, Moscow 117259, Russia\\     
     
\indent $^2$ \> Petersburg Nuclear Physics Institute, 188350,     
Gatchina, St. Petersburg, Russia. \\     

\indent $^3$ \> Univ. of Delaware, Bartol Res.\ Inst., Newark DE 19716,
USA.
     
\end{tabbing}     
     
\vspace*{2cm}     
     
\begin{abstract}     
Low $Q^2$ photon-proton cross sections
are analysed using a simple, QCD-motivated parametrisation 
$\sigma_{\gamma^\star p}\propto 1/(Q^2+Q_0^2)$, which
gives a good description of the 
data.   The $Q^2$ dependence of 
the $\gamma^\star p$ cross section is
discussed in terms of the partonic transverse momenta of the hadronic
state the photon fluctuates into.
\end{abstract}     
     
\newpage     


\section{Introduction}     
     
Traditionally photoproduction and DIS are considered as
processes which are governed by different underlying physics.
Whereas most of the features of DIS processes can be described in terms
of perturbative QCD, photoproduction is dominated by non-perturbative
effects.
This point of view seems to be justified by 
the $Q^2$ dependence of the $\gamma^\star p$ cross
section which exhibits a
clearly visible transition region between photoproduction and DIS at
about $Q^2 \sim 0.5 {\rm GeV}^2$. 
On the other hand, the steady transition from photoproduction to DIS 
highlights the importance of obtaining a 
description which smoothly links the non-perturbative and     
perturbative domains, see for example
\cite{Levy96a,Capella94b,Schildknecht97a,Desgrolard98a,Landshoff98a,Povh98a}.     

There exist now high precision deep inelastic lepton scattering     
data \cite{Aid96b,Adloff97a,Breitweg97b,Adams96a,Arneodo89a} 
covering both the low $Q^2$ and high $Q^2$     
domains, as well as measurements of the photoproduction cross     
section \cite{Derrick92a,Aid95b,Landolt87}.
In the present paper we discuss a simple QCD-motivated parametrisation
of the observed $Q^2$ dependence of the  $\gamma^\star p$ cross
section, which is closely related the average
transverse momentum of secondary particles produced in the photon
hemisphere.
In addition, the question of the
hard scale in deep-inelastic scattering is discussed within the
framework of $k_T$ factorization.


\section{Theoretical framework of $k_t$ factorization}     


Let $\sigma_{\gamma^\star p}(s, Q^2)$ be the     
total cross section for the process $\gamma^* p \rightarrow X$     
where $Q^2$ is the virtuality of the photon and $\sqrt{s}$ is the     
$\gamma^* p$ centre-of-mass energy. 
%
%
For $s\gg Q^2$ the
$\gamma^* \rightarrow
q\bar{q}$ fluctuations occur over a much longer time
scale than the interaction of the
$q\bar{q}$ pair with the target proton. Therefore 
the $\gamma^\star p$ 
cross section is well approximated by
the probability $|{\mathcal M}|^2$ of the $\gamma^* \rightarrow q\bar{q}$ 
transition multiplied by the    
imaginary part of the forward amplitude describing the $q\bar{q}$-proton     
interaction      
\be      
\Im m\left( A_{q\bar{q} + p}\right) \; = \; s \sigma_{q\bar{q} + p}\ ,
\label{eq:b12}      
\ee      
where $\sigma_{q \bar{q}+p}$ is the cross section for the
scattering of the $q\bar{q}$ system on the proton.
For transversely polarized photons the amplitude of the     
$\gamma^* \rightarrow q\bar{q}$ transition reads
\begin{equation}        
{\mathcal M}_T = \frac{\sqrt{z(1-z)}}{\bar{Q}^2 + k^2_T}     
\         
\bar{u}_{\lambda}(\gamma .     
\epsilon_{\pm})u_{\lambda^\prime}        
= \frac{(\epsilon_{\pm}.k_T)[(1-2z)\lambda \pm 1]         
\delta_{\lambda,-\lambda^\prime}        
+ \lambda m_q \delta_{\lambda\lambda^\prime}}         
{\bar{Q}^2 + k_T^2},      
\label{eq:a13}        
\end{equation}        
where the $q$ and $\bar{q}$ longitudinal momentum     
fractions and         
transverse momenta are $z,\; \vec{k}_t$ and $(1-z),\; -\vec{k}_t$ 
respectively.  
We use the notation of Ref.~\cite{Levin97a}, which is based     
on the earlier work of        
Ref.~\cite{Mueller90a,Brodsky80a}. Namely $\bar{Q}^2$ and the photon     
polarization vectors are         
given by        
\begin{eqnarray}        
\label{eq:a14}      
\bar{Q}^2 = z(1-z)Q^2+m^2_q  \\        
\epsilon_T = \epsilon_{\pm} = \frac{1}{\sqrt{2}}     
(0,0,1,\pm i),        
\label{eq:a15}      
\end{eqnarray}        
and where $\lambda, \lambda^{\prime} = \pm 1$ according     
to whether        
the $q, \bar{q}$ helicities are $\pm \frac{1}{2}$.        
        
In terms of the quark momentum variables we thus obtain      
\begin{equation}        
\sigma_{\gamma_T^\star p}(s,Q^2) = \sum_q \alpha \frac{e^2_q}{2\pi}     
\int dz\ dk^2_T \         
\frac{[z^2 + (1-z)^2]k^2_T+m^2_q}{(\bar{Q}^2 + k^2_T)^2}     
\         
N_c \sigma_{q\bar{q}+p} (s, k^2_T)        
\label{eq:a16}      
\end{equation}        
where the number of colours $N_c = 3$.

Eq.~(\ref{eq:a16}) can be rewritten as a dispersion relation in
$M^2$, with $M$ being the invariant mass of the $q\bar q$ pair.
With
\begin{equation}        
M^2 = \frac{k^2_T + m^2_q}{z(1-z)}      
\label{eq:a17}        
\end{equation}        
and a change of the integration variable from $dk_T^2$ to     
$dM^2$ one gets\footnote{
Of course, in principle, there may be non-diagonal elements of the
amplitude
$A_{q\bar{q} + p}$ of  (\ref{eq:b12}).  However it is known, both from
experiment  and from triple Regge theory, that such non-diagonal
transitions
are  suppressed in the forward direction.  In terms of the additive
quark model the suppression is the result of the orthogonality of the
initial and final wave  functions for a non-diagonal transition.}
\be      
\sigma_{\gamma_T^\star p}(s,Q^2) \; = \; \frac{\alpha}{2\pi} \sum_q     
e_q^2 \int dz \frac{dM^2}{(Q^2 + M^2)^2} \: \left \{ M^2 \left [z^2 +     
(1 - z)^2 \right ] + 2 m_q^2 \right \} \; N_c \sigma_{q\bar{q} + p}(s, k_T^2).
\label{eq:b17}      
\ee      
This can be compared with the corresponding expression of the
generalized vector dominance model
\cite{Sakurai72a,Sakurai72b,Donnachie78-b}
\begin{equation}
\sigma_{\gamma_T^*p}(s,Q^2) = \sum_q \int^{\infty}_0
\frac{dM^2}{(Q^2+M^2)^2}
\ \rho(M^2) \sigma_{q \bar{q}+p}(s,M^2),
\label{eq:a12}
\end{equation}
where the spectral function
$\rho$ represents the density of $q\bar{q}$ states.
A similar dispersion relation has been used, for example,
in \cite{Badelek92,Kwiecinski89a} 
to describe the structure function $F_2$ over the full $Q^2$ range.
In comparison to (\ref{eq:a12}) we see that     
(\ref{eq:b17}) is a two-dimensional       
integral.  To see the reason for this let us consider massless quarks.      
Then $z = \frac{1}{2} (1 + \cos       
\theta)$ where $\theta$ is the angle between the $q$ and     
the $\gamma^*$ in the       
$q\bar{q}$ rest frame.  The $dz$ integration is implicit     
in (\ref{eq:a12}) as the       
integration over the quark angular distribution in the     
spectral function $\rho$.

At first sight the $Q^2$ dependence of the cross section (\ref{eq:a12}) should 
be $\sigma_{\gamma^* p}\propto 1/(Q^2+M_0^2)^2$. 
This is true if one deals with only one vector meson or when the 
dominant contribution in (\ref{eq:a12}) 
comes from a limited range of $M^2$. 
On the other hand when all the possible values of $M^2$ are taken into 
account the result is 
\be
\sigma_{\gamma^* p}\propto \int\frac{dM^2}{(Q^2+M^2)^2}
=\frac 1{Q^2+M_0^2}\; .
\ee
Just such a behaviour is expected in our approach 
(see Sect.~3 and Eq.~(\ref{eq:b20})).

To obtain a complete description of the $\gamma^\star p$ cross section a
deed model for the $q\bar q$-proton interaction is needed. Such a
model is developed, for example, in \cite{Martin98a}. Furthermore,
longitudinally polarized photons have to be considered. 
However for our phenomenological
discussion it is sufficient to investigate some general features of
(\ref{eq:a16}).
        

\section{Virtuality dependence}

The $Q^2$ dependence in Eq.~(\ref{eq:a16}) comes mainly from the two 
quark propagators $1/(\bar{Q}^2 + k^2_T)^2 = 1/(z(1-z)Q^2+k^2_T)^2$
where in the r.h.s. of the last equality we neglect the small quark 
mass ($m_q^2$) in the $\bar{Q}^2$ term. In order to demonstrate the 
expected $Q^2$ behaviour of the cross section (\ref{eq:a16}) let us 
first write the expression in the simplified form
\be
\sigma_{\gamma_T^\star p}(Q^2) \propto \int^{1/2}_0\frac{dz}{(zQ^2+k^2_T)^2}
\ee
and perform the $dz$ integration. It gives the result
\be
\sigma_{\gamma_T^\star p}(Q^2) \propto \frac 1{k^2_T(Q^2+2k^2_T)}\propto \frac 
1{Q^2+2k^2_T}.
\ee 
It can be checked by explicit numerical integration that the $z$
dependent part of the integral (\ref{eq:a16})
\be
J_\sigma=\int^1_0 dz\frac{[z^2+(1-z)^2]}{(z(1-z)Q^2+k^2_T)^2}
\ee
is well approximated by
\be
J_{\rm app.}=\frac{2}{k^2_T(Q^2+3k^2_T)}\ .
\ee
The ratio $r=J_\sigma/J_{\rm app.}$ tends
to 1 in the asymptotic limits $Q^2\to 0$ or $Q^2\to \infty$ and reaches a 
minimum of about 0.96 at $Q^2\sim 8k^2_T$. 

Using this approximation (\ref{eq:a16}) can be written as
\bea
\sigma_{\gamma_T^\star p}(s,Q^2) &=& N_c\alpha\sum_q
\frac{e_q^2}{2\pi}\int d\log(k_T^2) \frac{2}{Q^2+3 k_T^2}
k_T^2\sigma_{q\bar{q}+p}(s,k_T^2)
\nonumber\\
&\propto & N_c\alpha\sum_q
\frac{e_q^2}{2\pi} \frac{2}{Q^2+3 \overline{k_T^2}}
\overline{k_T^2} \sigma_{q\bar{q}+p}(s,\overline{k_T^2})\ ,
\label{eq:b19}
\eea
where $\overline{k_T^2}$ is the characteristic transverse momentum of
the quark.
In Eq.~(\ref{eq:b19})
the $Q^2$ dependence is almost factorised and is mainly given 
just by the factor $1/(Q^2+3 \overline{k_T^2})$.


Now let us discuss the structure of the integral (\ref{eq:b19}) over $dk^2_T$. 
Of course the large distances, i.e. very small $k_T<1/r$ (where $r\sim R_N$ is
 of the order of nucleon radius $R_N$) are suppressed by the confinement. At
 very large $k_T\gg 1/r$ based on the perturbative QCD and neglecting the 
anomalous dimension one expects the cross section 
$\sigma_{q\bar{q}+p}\propto 1/k^2_T$. 
Thus in the ultraviolet region 
(at $k^2_T>Q^2$) the integral (\ref{eq:b19}) is convergent.

The main contribution comes from the mediate (more or less soft) 
$k^2_T\sim 0.1\, -\, 0.4$ GeV$^2$ interval. 
Typically the cross section is large in the soft region, 
where it falls down with $k_T$ exponentially; 
then at $k_T>1\, -\, 3$ GeV 
(corresponding to a small distances) it has the power-like tail .

Note that the predicted behaviour 
\be
\sigma_{\gamma^* p}\propto \frac 1{Q^2+3\overline{k^2_T}}
\label{eq:c40}
\ee
does not depend 
too much on concrete form of the $q\bar{q}$ cross 
section.

As an  extreme example let us consider a simple "soft" approximation
\be
\sigma_{q\bar{q}+p}(s,k_T^2) = \sigma_0(s) \Theta(k_T^2-\mu^2)
\Theta(\overline{K_T^2}-k_T^2)
\ee
which corresponds to soft scattering where the quark-proton cross
section is saturated for $\mu^2 < k_T^2 < \overline{K_T^2}$ and vanishes
everywhere else.
Then the $\gamma^\star p$ cross section reads
\be
\sigma_{\gamma_T^\star p}(s,Q^2) 
\propto
\sigma_0(s) \ln \left(\frac{Q^2+3 \overline{K_T^2}}{Q^2+3 \mu^2}\right)\
.
\label{eq:b21}
\ee
Despite of the fact that (\ref{eq:b21}) takes now a logarithmic form 
for the numerical values for $\overline{K_T^2}$ discussed in the following
(\ref{eq:b21}) predicts almost the same $Q^2$ behaviour (\ref{eq:c40}) as 
Eq.~(\ref{eq:b19}).
 

One has to expect also that the characteristic value 
$\overline{k^2_T}$ should increases with energy. 
For larger collision energies the evolution chain 
(which produces finally the wee parton) becomes longer. 
At each step of evolution a new parton is emitted
and the active parton gets some 
extra transverse momentum. Therefore, as in the case of multiple small angle 
scattering in a thick target, the final 'intrinsic' $k_T$ of the active 
parton grows 
with the number of interactions (the number of evolution steps).
In the framework  of perturbative QCD  this growth is originated in the 
summation of the
 double logarithmic contributions of the type 
$(\alpha_s\log{(k^2_T)}\log{(s)})^n$ and 
 is described in terms of the
 anomalous dimensions.  Due to the larger
 value of anomalous dimension $\gamma$ 
 at higher energies one expects
the larger characteristic value $\overline{k_T^2}$. 

Finally 
we will neglect the weak logarithmic $Q^2$ dependence of
$\overline{k_T^2}$ in (\ref{eq:b19},\ref{eq:c40}) 
(which is, of course, not excluded completely)  
and try to describe the data with the parametrisation
\be
\sigma_T(\gamma^*p) \propto \frac 1{Q^2+Q^2_0}
\label{eq:b20}
\ee
using $Q_0^2$ given by the characteristic value $\overline{k^2_T}$ of the 
quark transverse momentum
\be
Q_0^2 \approx
3 \overline{k^2_T}\ .
\label{eq:c1}
\ee
The value of $Q^2_0$  becomes unimportant for large $Q^2$ so we
use the value of $\overline{k^2_T}$ as determined at small $Q^2$ (say, 
in photoproduction at $Q^2=0$).

Since the integral (\ref{eq:b19}) over $k^2_T$ has a logarithmic structure 
one can not estimate the characteristic value of $\overline{k_T^2}$ through
the average of $k^2_T$. 
Multiplying the integrand by an extra power of 
$k^2_T$ destroys the whole structure of the integral and enlarges 
crucially the essential values of $k^2_T$. Therefore we estimate
$\overline{k_T^2}$ by averaging the logarithm of the squared transverse
momentum 
\be
\overline{k^2_T} =  \exp\left(
\langle\log(k^2_T)\rangle\right)\ .
\ee

Of course, one cannot  measure directly $k^2_T$ of a quark. 
So we have to relate the $k_T$ of the quark jet to the transverse 
momenta $p_T$ of secondary hadrons in the photon fragmentation
region.
Contrary to the large $E_T$ jet fragmentation here 
(for not too large $k_T$)  the value of $p_T$ is not so small in comparison 
with $k_T$. If one assumes that in photoproduction both values ($k_T$ of 
the quark and $p_T$ of secondary hadrons) are controlled by the typical
temperature $T$ then we may expect that our $\overline{k^2_T}$ is close
(or equal) to the average $p^2_T$ 
of secondary hadrons (in the photon 
fragmentation region).

To understand better the relation between the values of $\overline{k^2_T}$ and
 $\overline{p^2_T}$ we use a standard Monte Carlo program which is in agreement 
 with fixed target 
 and HERA photoproduction data, in particular with the measured transverse 
 is momentum spectra of secondaries.
%
The corresponding predictions for the {\sc Phojet}  
(which, of course is not excluded completely)  
event generator \cite{Engel95d} are given in Tab.~\ref{tab:1} for two
energies at which inclusive transverse momentum distributions of
secondaries have been measured \cite{Apsimon89a,Abt94a,Derrick95i}.
Indeed for the photon fragmentation region 
($x_F>0.2$) the $\overline{p^2_T}$ of secondary hadrons 
in non-diffractive events 
is close to the parton $\overline{k^2_T}$ and increases with energy.
A similar increase of the $\overline{p^2_T}$ of secondary hadrons with the
collision energy was observed experimentally in 
deep inelastic scattering~\cite{Derrick96f,Ashman91a,Adams91a}.

\begin{table}[htb]
\caption{\label{tab:1}
Logarithmic average transverse momenta 
of partons ($k_T$) and charged final state
hadrons ($p_T$) produced in non-diffractive $\gamma p$ collisions 
in a photon fragmentation ($x_F>0.2$) region as predicted by 
the {\sc Phojet} event generator \cite{Engel95d}. 
In the last column the $Q_0^2$ values as obtained by a fit to 
photoproduction and DIS data are given.
}
\renewcommand{\arraystretch}{1.3}
\begin{center}
\begin{tabular}{|c|c|c|c|}\hline
$\sqrt{s}$~~(GeV) &  $\overline{k^2_T}$~~(GeV$^2$) &
$\overline{p_T^2}$~~(GeV$^2$) & $Q_0^2$~~(GeV$^2$)
\\ \hline
15  &  0.19  & 0.17 &    $0.42\pm 0.01$
\\ \hline
200 &  0.5   & 0.35  &    $1.04\pm 0.04$
\\ \hline
\end{tabular}
\end{center}
\end{table}

In Fig.~1 the parametrisation (\ref{eq:b20}) is compared to
photoproduction~\cite{Aid95b,Landolt87} and 
DIS~\cite{Aid96b,Adloff97a,Breitweg97b}
data at two different energies ($\sqrt{s}= 200$~GeV and $\sqrt{s}=
15$~GeV). To obtain the total virtual photon--proton cross sections
we use $\sigma_{\gamma^*p}=(4\pi\alpha/Q^2)F_2(x,Q^2)$.
Where necessary, the measured values
of $F_2(x,Q^2)$ have been interpolated using a smooth function.
The overall normalization uncertainties have been neglected in the
fit of the data in the Fig.~1.

The measured total photon--proton cross sections are fitted
to the parametrisation (\ref{eq:b20}) with only two free parameters:
$Q_0^2$ and an overall normalization for each of the two sets of the
data. The results of the fit are also presented in Fig.~1 as lines.
We can conclude that (\ref{eq:b19},\ref{eq:b20}) reproduce all the main 
features of
the data (including the photoproduction points) in a wide range of 
energies and $Q^2$.
As expected, the parameter $Q^2_0$ is energy dependent. It's value is 
$1.04\pm 0.04$~GeV$^2$ at $\sqrt{s}=200$~GeV while for $\sqrt{s}=15$~GeV we
needed  $Q^2_0=0.42\pm 0.01$~GeV$^2$.
Relating the experimentally measured transverse momentum spectrum 
to the parameter $\overline{k_T^2}$, the increase of $\overline{k_T^2}$ 
from the energies of fixed target experiments to the
HERA kinematic region agrees well the rise of $Q^2_0$ parameter obtained
above (see Eq.~(\ref{eq:c1})).

It is known that in the small-$x$ region the DGLAP evolution leads to a strong 
scaling violation which reveals itself in a rather large positive value of the 
anomalous dimension.
Therefore, at large energies the $Q^2$ behaviour of the 
cross section $\sigma\propto 1/(Q^2)^{1-\gamma}$ becomes more flat.
In our parametrisation (\ref{eq:b20}) the same effect is hidden in the value
 of $Q^2_0\propto k^2_T$. Due to a rather large anomalous dimension
and $\sigma_{q\bar{q}+p}(s,k^2_T)\propto 1/(k^2_T)^{1-\gamma}$,
the essential values of $k^2_T$ increase with energy faster than 
$\log(s)$. Consequently, the $Q^2$ distribution becomes broader.

The expression (\ref{eq:b20}) fits also quite well the data on the nuclear 
targets. 
In Fig.~2 the parametrisation (\ref{eq:b20}) is compared to the 
available photoproduction~\cite{Landolt87} and
DIS~\cite{Adams96a,Arneodo89a} data at energy $\sqrt{s}= 10$~GeV for the nucleon
in deuteron, carbon and calcium nuclei.
The value of $Q^2_0$ increases 
with the atomic number $A$ but the normalization factors are the same
within the errors. 
In  other words not only the anomalous dimension behaviour of  DIS cross 
section but the effect of shadowing is absorbed in the value of $Q^2_0$ 
(to be more precise -- in the $A$-dependence of $Q^2_0$) as well.

What is the origin of $A$-dependence of $Q^2_0$?
{}From the point of view of the photon-quark interaction 
the $k_T$ in Eq.~(\ref{eq:b19}) 
 plays the role of the intrinsic transverse momentum of the quarks. So we have 
to discuss the parton wave function of the nucleon/nuclear target. 
Note that in coordinate space the longitudinal interval occupied by
 the small-$x$ wee parton $z\sim 1/m_Nx$ ($m_N$ is the nucleon mass) increases
 with $1/x$ and for $x< 0.1\, - \, 0.2$ the partons originated by 
different nucleons start to overlap and interact with each other. 
 This parton-parton rescattering leads to the well-known shadowing effects.
 However a quark can not disappear completely since it carries conserved
quantum numbers (i.e.\
 charge, isospin, etc.). The only possibility is to move the quark from the 
densely populated phase space region to another one. 
Consequently the rescattering mainly enlarges 
the transverse momentum and pushes the quark out of the small $k_T$ region.

Therefore we have to expect 
a larger value of 
$Q_0^2=3\overline{k^2_T}$ for a heavier nuclei. To estimate the size of 
this effect let us consider soft rescattering. With a quark-nucleon cross 
section of $\sigma_{qN}\simeq \frac 13\sigma_{NN}\simeq 10\, -\, 15$ mb 
the mean number of quark interactions in $Ca$ will be $\nu_q\sim 0.7\, -\, 1$.
 Each soft rescattering increases the value of $\overline{k^2_T}$ by 
about $\Delta k^2\sim (0.3-0.4$~GeV$)^2$ since
$k^2_A \approx \overline{k^2_T}+\nu_q\cdot\Delta k^2$. 
The parameter $Q^2_0$ for  $Ca$ should be of
about $3\nu_q\cdot\Delta k^2\sim 0.3-0.4$GeV$^2$ larger 
than for a free nucleon target. This is in good agreement with the fit
results given in Tab.~\ref{tab:2}.
\begin{table}[htb]
\caption{\label{tab:2}
Fit results for the parameter $Q_0^2$ for different energies and
targets. The data sets have been interpolated to the given energy.
}
\renewcommand{\arraystretch}{1.3}
\begin{center}
\begin{tabular}{|c|c|c|c|}\hline
$\sqrt{s}$~~(GeV) &  target & $Q_0^2$~~(GeV$^2$) & data used
\\ \hline
200 &  p     &  $1.04\pm 0.04$ & H1, ZEUS
\\ \hline
100 &  p     &  $0.75\pm 0.04$ & H1, ZEUS
\\ \hline
15  &  p     &  $0.42\pm 0.01$ & E665
\\ \hline
10  &  d     &  $0.46\pm 0.02$ & E665
\\ \hline
10  &  d     &  $0.44 \pm 0.03$ & EMC
\\ \hline
10  &  C     &  $0.56 \pm 0.06$ & EMC
\\ \hline
10  &  Ca    &  $0.76 \pm 0.08$ & EMC
\\ \hline
\end{tabular}
\end{center}
\end{table}

The same parton-parton rescattering may explain at least part of the growth of
 $Q^2_0$ with the energy; due to the fact that at large energies (small $x$) 
the parton density increases and even in the case of a single proton the  
parton-parton interaction becomes not negligible.

In terms of the dispersion relation (\ref{eq:a12}) 
one can say that in dense matter (heavy nuclei or large $\sqrt{s}$) the 
effective mass ($M^2$) of 
 $q\bar{q}$-pair increases. The point-like (large $M^2$) configurations 
with a small cross section which penetrate a thin target 
without any interaction are 
absorbed by a dense target and give an essential contribution to the 
cross section.


\section{Conclusive remarks}

At low $x$ photon--proton scattering can be understood as the fluctuation 
of the virtual photon into a hadronic state and the subsequent scattering of
this state on the proton. 
We have shown that cross section data on fixed target and HERA deep-inelastic
scattering support this interpretation. 

On this basis, a simple
parametrisation of the $Q^2$ dependence of the $\gamma^\star p$ cross
section has been derived. The essential parameter $Q_0^2$ of this
parametrisation can be estimated from the measurement of secondaries
produced in the photon fragmentation region.

This data analysis also confirms the prediction of the $k_T$
factorization approximation that the hard scale of the process 
is not the initial photon virtuality $Q^2$ but the parton $k_T$ of the 
hadronic fluctuation. Of course, the essential values of $k^2_T$ are 
correlated with $Q^2$ but neither directly equal nor proportional to $Q^2$.
Instead, the correlation between $k_T^2$ and $Q^2$ is rather broad.
Therefore, in order to fix the hard scale of 
the deep-inelastic process it is better to use the transverse energy
($E_T$) measurements in the photon fragmentation region, than the
value of the colliding photon virtuality.
Fixing the hard scale by high $p_T$
secondary hadrons from the photon fragmentation region instead of the
photon virtuality $Q^2$, a faster transition 
to hard scattering has been observed \cite{Aid97b}.


Finally, we may say that 
the cross section fits presented in this work suggest that low-$x$ deep
inelastic scattering is 
characterized by rather "soft" (corresponding to 
$k^2_T\sim 0.15\, -\, 0.3$ GeV$^2$) quark-nucleon 
($\sigma_{q\bar{q}+p}$) interactions.

\clearpage
 
\noindent {\large \bf Acknowledgements}        
        
MGR thanks the INTAS (95-311) and the Russian Fund of        
Fundamental Research (98-02-17629) for support. One of the authors
(RE) is supported by the U.S.\ Department
of Energy under grant DE-FG02-91ER40626.
AR gratefully acknowledge the University of Antwerpen for support.          
\newpage        
        



\newpage        
        
\noindent {\large \bf Figure Captions.}        

{\bf Figure 1:}
Total $\gamma{p}$ cross section as function of $Q^2+Q_0^2$.
The filled circles represent HERA deep inelastic data at 
$W=200$~GeV, triangles show the data from E665 experiment
at $W=15$~GeV, the squares represent the photoproduction
measurements at corresponding energies.

{\bf Figure 2:}
Total cross section of photon--nucleon interaction
as function of $Q^2+Q_0^2$ at $W=10$~GeV.
The filled circles represent EMC data 
on $\mu d$, $\mu C$, and $\mu Ca$ deep inelastic scattering,  
triangles show the $\mu d$ data from E665 experiment, the squares
represent the photoproduction $\gamma d$ measurement. 

\newpage
 
\begin{figure}[h] \unitlength 1mm
 \begin{center}
   \begin{picture}(50,100)
    \put(-35,-70){\epsfig{file=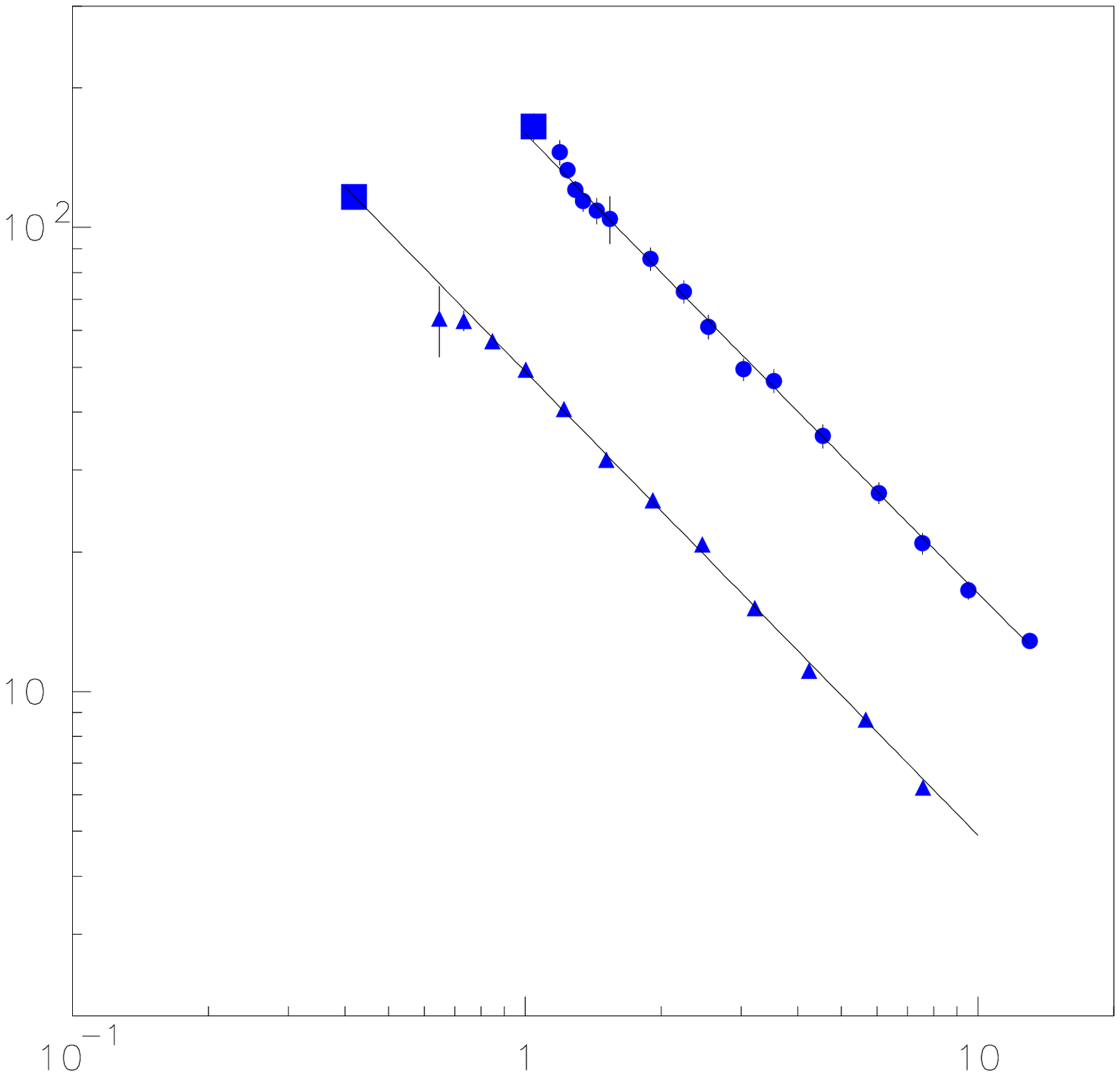,bburx=500,
        bbury=543,bbllx=69,bblly=90,height=12.0cm}}
    \put(-60,50){\bf\Large $\sigma_{\gamma^\star{p}}/\mu{b}$}
    \put(40,-53){\bf\Large $(Q^2+Q_0^2)/GeV^2$}
   \end{picture}
 \end{center}
\end{figure}
\vspace*{5.0cm}
\begin{center}
{\bf\Large Figure 1}
\end{center}
\newpage
\begin{figure}[h] \unitlength 1mm
 \begin{center}
   \begin{picture}(50,100)
    \put(-35,-70){\epsfig{file=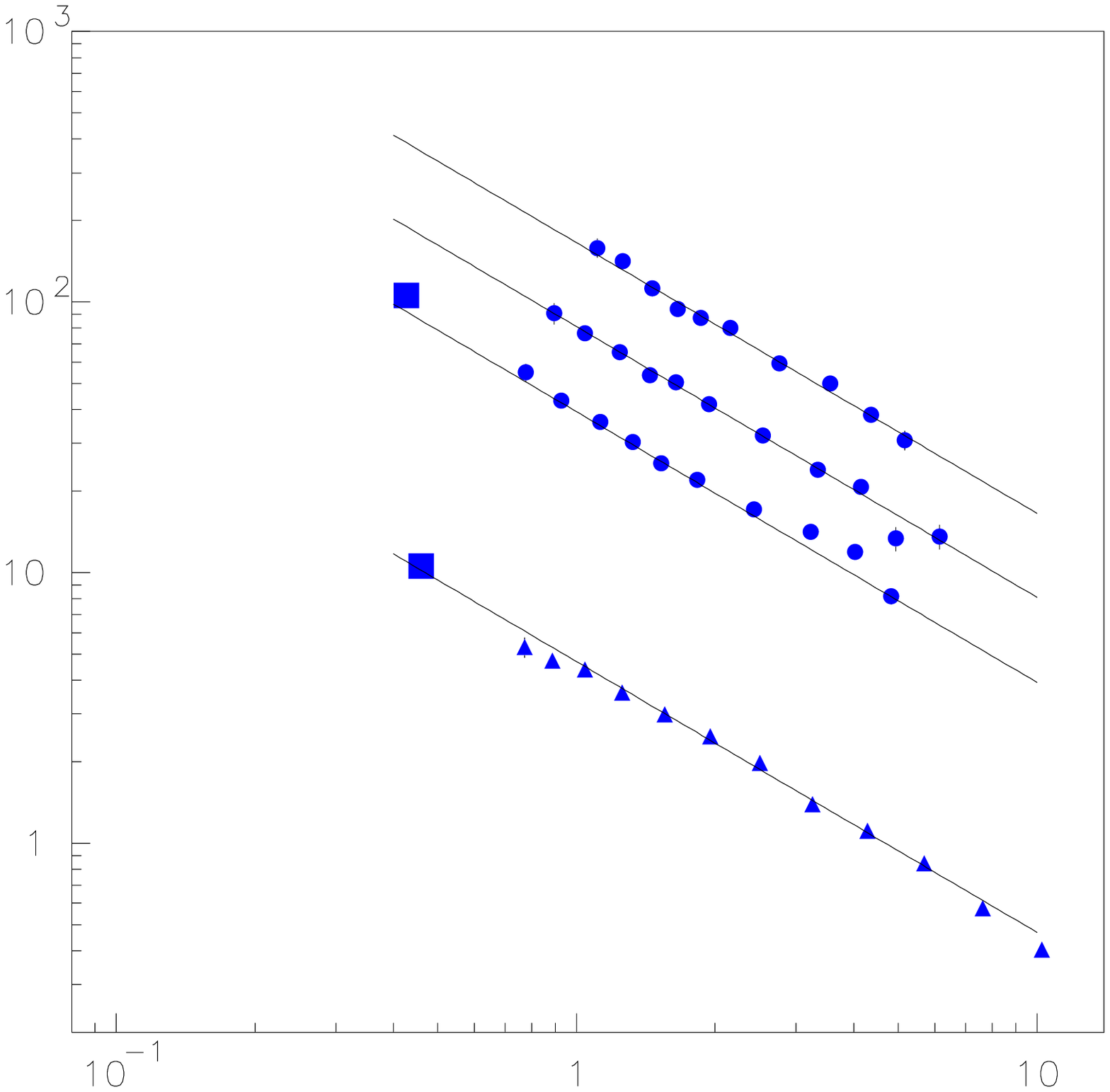,bburx=500,
        bbury=543,bbllx=69,bblly=90,height=12.0cm}}
    \put(-20,65){\bf\Large $Ca~(\times 4)$}
    \put(-18,55){\bf\Large $C~(\times 2)$}
    \put(-18,45){\bf\Large $d~(\times 1)$}
    \put(-20,15){\bf\Large $d~(\times 0.1)$}
    \put(-60,50){\bf\Large $\sigma_{\gamma^\star{N}}/\mu{b}$}
    \put(40,-53){\bf\Large $(Q^2+Q_0^2)/GeV^2$}
   \end{picture}
 \end{center}
\end{figure}
\vspace*{5.0cm}
\begin{center}
{\bf\Large Figure 2}
\end{center}
     
\end{document}